\documentclass[
	aps
	,pra
	,twocolumn
	,amsfonts
	,amssymb
	,amsmath
	,a4paper
	,showpacs
	,10pt
	,floatfix
]{revtex4}

\usepackage{amsmath}
\usepackage{amssymb}

\usepackage{dsfont}

\usepackage[T1]{fontenc}
\newcommand{\flqq}{\guillemotleft}
\newcommand{\frqq}{\guillemotright}

\usepackage{graphicx}

\newcommand{\ket}[1]{|#1 \rangle}
\newcommand{\ketbra}[1]{| #1 \rangle \langle #1|}
\newcommand{\average}[1]{\langle #1 \rangle}
\newcommand{\Phip}{\mathbf{\Phi}^{+}}
\newcommand{\Phim}{\mathbf{\Phi}^{-}}
\newcommand{\Psip}{\mathbf{\Psi}^{+}}
\newcommand{\Psim}{\mathbf{\Psi}^{-}}

\newcommand{\rhoh}{\hat{\rho}}
\newcommand{\rhos}{\rhoh^\text{(s)}}
\newcommand{\rhou}{\rhoh^\text{(u)}}
\newcommand{\probs}{p^\text{(s)}}
\newcommand{\fidphip}{F}
\newcommand{\fidu}{F^\text{(u)}}
\newcommand{\As}{A^\text{(s)}}
\newcommand{\Bs}{B^\text{(s)}}
\newcommand{\Cs}{C^\text{(s)}}
\newcommand{\Ds}{D^\text{(s)}}
\newcommand{\Fit}{F_\text{it}}

\begin{document}

\title{Iterative entanglement distillation for finite resources}
\author{Stefan Probst-Schendzielorz} 
\email[E-mail me at: ]{Stefan.Probst@uni-ulm.de}

\author{Thorsten Bschorr} 

\author{Matthias Freyberger}

\affiliation{Abteilung Quantenphysik, Universit\"at Ulm, 89069 Ulm, Germany}

\begin{abstract}
We discuss a specific entanglement distillation scheme under the constraint of finite samples of entangled qubit pairs. It is shown that an iterative process can be explicitly formulated. The average fidelity of this process can be enhanced by introducing conditional storing of entangled qubit pairs in each step of the iteration. We investigate the corresponding limitations on the size and the initial fidelity of the sample. 
\end{abstract}


\maketitle

\section{Introduction}

The quantum concept of entanglement is the most intriguing feature that allows to establish new physical paradigms in information processing. The corresponding  non-classical protocols need a certain degree of entanglement as a quantum resource \cite{BEN98}. Consequently one has to understand how entanglement can be processed and how it can be measured \cite{VEP98,MHO01}.

In this respect the typical situation is the following. Several parties share components of an entangled system. Processing of entanglement then means that they can perform local unitary operations and local measurements on their respective parts of the complete system. Under these conditions it has been shown \cite{BBP96, BDS96, DEM96, GIS96, DNMV03} that two parties, supplied with non-maximally entangled qubit pairs, can extract a sample of stronger entangled pairs. 
But so far no general approach to optimal distillation exists. However, in order to improve the fidelity, modifications of the original versions have been proposed \cite{OKU99,MS02,MET02}. Lower bounds for the fidelity of entanglement distillation based on faulty local operations have also been studied \cite{GBCZ99}.

In the present work we investigate the entanglement distillation protocol described in \cite{DEM96} for the case that only a finite sample of entangled qubit pairs is available. In particular, we propose a relatively simple, iterative distillation scheme that starts from a finite number of identical pairs and delivers a distilled pair applicable for further communication tasks. The behaviour of the corresponding mean fidelity turns out to be particularly interesting for small initial samples of entangled qubit pairs.

\section{Entanglement distillation}

For any non-classical communication, entangled systems first have to be distributed between a sender (Alice) and a receiver (Bob). In the course of this distribution the systems are influenced by various noise sources, which reduce the amount of entanglement. Any distillation process for such distributed systems is restricted to local operations and classical communication (LOCC) of the parties. Moreover, realistic entanglement distillation schemes have to take into account errors \cite{GBCZ99} and the fact that Alice and Bob share only a finite amount $N$ of mixed entangled systems. In the present paper we focus on the latter restriction regarding resources.

The explicit distillation protocol we refer to was introduced in Refs. \cite{BBP96,DEM96}. In the present work we only study this specific distillation process under the constraint of finite resources. However, other highly effective distillation protocols have been proposed in the literature. Their exclusion in the present discussion does not mean that they cannot be applied to finite resources. In particular, the quantum hashing protocol \cite{BDS96} can be well applicable to finite resources and presumably leads to highly distilled qubit pairs. It will, however, need a considerable effort for the corresponding book keeping of the so-called likely sets.

The process \cite{BBP96,DEM96} conditionally increases the fidelity
\begin{equation}
 F(\rhoh) \equiv \operatorname{tr}\left(\Phip \rhoh \right) = A
\end{equation}
of the Bell-diagonal mixed state
\begin{equation}
\label{belldiag}
\rhoh  = A \: \Phip \: + B \: \Psim \: + C \: \Psip \: + D \: \Phim \;
\end{equation}
describing a qubit pair, where we introduced the abbreviations $\mathbf{\Phi}^{\pm} \equiv \ketbra{\phi^\pm}$ and $\mathbf{\Psi}^{\pm} \equiv \ketbra{\psi^\pm}$ for the four Bell-states. 

This state is non-separable if any of the coefficients $A$, $B$, $C$ or $D$ is larger than $1/2$ \cite{MPRH97}. Without loss of generality we choose $A>1/2$.

The central element of the protocol is a CNOT transformation. We briefly recall the basic ideas \cite{BBP96,DEM96}. Two qubit pairs 1 and 2 each described by the state $\rhoh$, Eq.~\eqref{belldiag}, are processed in one step. Alice holds the qubits $1_A$ and $2_A$ and Bob holds the qubits $1_B$ and $2_B$. These qubits can now be treated locally using a sequence of operations \cite{DEM96}: (I) Alice and Bob rotate their qubits locally. These rotations exchange the $\Psim$ contribution with the $\Phim$ contribution of the initial state $\rhoh$, Eq.~\eqref{belldiag} \footnote{Note that this step is necessary for an iterative application of the distillation scheme since otherwise it does not converge, for more details see the analysis in \cite{CMA98}.}. (II) Alice and Bob then perform CNOT operations on their respective qubits. The qubits of pair 1 (qubits $1_A$ and $1_B$) act as control qubits. (III) Both measure their target qubits (qubits $2_A$ and $2_B$) in the computational basis $\{\ket{0},\ket{1}\}$. They obtain either the result {\frqq{$0$}\flqq} or {\frqq{$1$}\flqq}. (IV) Alice and Bob classically communicate their measurement results. The distillation is successful with probability
\begin{equation}
\label{ps}
\probs \equiv (A+B)^2+(C+D)^2
\end{equation}
if the combined result reads {\frqq{$00$}\flqq} or {\frqq{$11$}\flqq}. Then they keep pair 1 now described by the conditioned density operator
\begin{align}
\label{rhosvier}
\begin{split}
\rhos & = \frac{1}{\probs}\bigg[ \;(A^2+B^2) \: \Phip \: + 2 C D \: \Psim\: \\
             & \phantom{=}  +\; (C^2+D^2) \: \Psip \: + 2 A B \: \Phim \;\bigg].
\end{split}
\end{align}
If Alice and Bob read off the results {\frqq{$01$}\flqq} or {\frqq{$10$}\flqq}, the reduced density operator of pair 1 reads
\begin{equation}
\label{rhouvier}
\begin{split}
\rhou & = \frac{1}{1-\probs} \bigg[ \; (AC+BD) \Phip + (AD+BC) \Psim \\
            & \phantom{=} +\; (AC+BD) \Psip + (AD+BC) \Phim\;\bigg].
\end{split}
\end{equation} 
In this unsuccessful case Alice and Bob discard pair 1.

In the successful case pair 1 is mapped from a Bell-diagonal state $\rhoh$, Eq.~\eqref{belldiag}, to the Bell-diagonal state $\rhos$, Eq.~\eqref{rhosvier}. The corresponding new fidelity is given by 
\begin{equation} 
\label{fsucc}
\fidphip(\rhos) = \frac{A^2+B^2}{(A + B)^2 + (C + D)^2}\;,
\end{equation}
which is higher than the initial fidelity for $A>0.5$. 

However, this fidelity does not completely describe the single-step process. A complete description also has to take into account the unsuccessful case. The distillation process then leads to the conditioned density operator $\rhou$, Eq.~\eqref{rhouvier}. Hence if we are interested in the performance of this particular distillation process, we have to take into account the fidelity $F(\rhou) \leq \frac{1}{2}$. On the other hand, using local operations and classical communication the parties can always generate two qubits with fidelity $\frac{1}{2}$ if the unsuccessful case occurs. Applying such additional operations we can define the average fidelity
\begin{equation}
\label{aveinzel}
\begin{split}
  \average{\fidphip}  & \equiv \probs \fidphip(\rhos) + (1-\probs) \frac{1}{2}\\
      & = A + B (1 - 2 A)
\end{split}
\end{equation}
for a single step process.
For all possible values of $B$ the average fidelity $\average{\fidphip}$ turns out to be equal to or less than the original fidelity $A$. This is of course also true for the average fidelity
\begin{equation}
\begin{split}
  \average{\tilde{\fidphip}}  & \equiv \probs \fidphip(\rhos) + (1-\probs) \fidphip(\rhou) \\
      & = A^2 + A(C-B) + B(1-C)
\end{split}
\end{equation}
based on the density operator $\rhou$, Eq.~\eqref{rhouvier}.
This is consistent with the fact that the total amount of entanglement cannot increase under a LOCC process \cite{MHO01}.

However, we emphasise that so far we have just discussed a single distillation step performed on two entangled qubit pairs. The situation becomes different and more interesting when we now consider a finite ensemble with $N>2$ pairs. 

\section{Distillation for a finite set of entangled systems}

Alice and Bob now perform the distillation scheme with an even number
$N$ of pairs\footnote{ Due to the fact that the entangled qubit pairs are being processed pairwise in the protocol, we assume an even number of pairs. For a single application of the protocol this assumption is without loss of generality.}.

After one distillation step they have $j \leq N/2 $ pairs left with probability 
\begin{equation}
\label{pn}
p(N,j) = \binom{N/2}{j} \big[\probs\big]^j\big[1-\probs\big]^{N/2-j},
\end{equation}
which contains the success probability $\probs$, Eq.~\eqref{ps}.

\begin{figure}[h!]
\begin{center}
\includegraphics{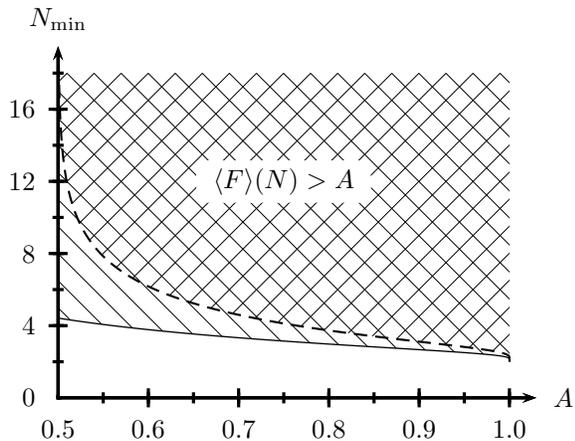}
\end{center}
\caption{\label{ensemblesize} Minimal size $N_\text{min}$, Eq.~\eqref{maxensemble}, of the finite sample of entangled qubits to obtain an average fidelity $\average{F}(N_\text{min})=A$ in a single step distillation. The curves have been calculated for specific Bell-diagonal states (Werner-states) with $B=C=D=(1-A)/3$. Nevertheless, they show the generic behaviour. The dashed curve results from the substitution $\fidu=\fidphip(\rhou)$ in Eq.~\eqref{maxensemble} and diverges for small initial fidelities. In contrast to this, the minimal size for the LOCC choice $\fidu=\frac{1}{2}$ stays always finite (solid curve). Clearly for $A\!\to\!1$ we obtain $N_\text{min}=2$.}
\end{figure}

The totally unsuccessful case occurs when all pairs have to be discarded, that is $j=0$, in a single distillation step. Hence starting from $N$ pairs we can define the average fidelity
\begin{equation}
\label{midfidelity}
\average{\fidphip}(N) \equiv p(N,0) \; \fidu \; + [1-p(N,0)] \; \fidphip(\rhos)
\end{equation}
using the density operators $\rhoh^{(s)}$, Eq.~\eqref{rhosvier}, and the fidelity $\fidu$ for the unsuccessful case.
With an increasing amount $N$ of pairs the average fidelity $\average{\fidphip}(N)$ also increases because of the decreasing probability $p(N,0)$ to lose all qubit pairs. Using Eq.~\eqref{midfidelity} one finds that
\begin{equation}
\label{maxensemble}
N_{\text{min}} = 2 \frac{\:\displaystyle \ln \left[ \frac{A - \fidphip(\rhos) }{ \fidu - \fidphip(\rhos) } \right] \:}{\displaystyle \ln(1-\probs)}
\end{equation}
pairs are needed to obtain an average fidelity $\average{\fidphip}(N_\text{min})=A$. This minimal number $N_\text{min}$ of qubit pairs shows a strong difference depending on the choice for $\fidu$. Figure \ref{ensemblesize} shows the minimal size $N_\text{min}$ for a finite ensemble described by Werner-states ($B=C=D=(1-A)/3$) \cite{WER89}. If we simply choose $\fidu=\fidphip(\rhou)$ to characterise the CNOT distillation itself we obtain a strong dependence (dashed curve) on the initial fidelity $A$. In particular, $N_\text{min}$ diverges like $\ln(A-\frac{1}{2})$ for $A\rightarrow\frac{1}{2}$. If, on the other hand, we substitute the LOCC boundary $\fidu=\frac{1}{2}$, we always get finite values for $N_\text{min}$ which only depend weakly on $A$. In the plot one can therefore clearly identify the region where on average the fidelity increases, that is $\average{F}(N)>A$.

Moreover, if $j \geq 2 $ pairs are left after such a first step we can continue with the distillation. We consider such an iteration in the following paragraph.

\section{Iterative distillation}

The resulting state $\rhos$ of a successful distillation step is again Bell-diagonal and hence the complete process can be applied iteratively, as long as qubit pairs are left to use. Starting from an initial density operator $\rhos_0 \equiv \rhoh$, Eq.~\eqref{belldiag}, the density operator $\rhos_{i-1}$ is mapped on a density operator $\rhos_i$ if the $i^\text{th}$ distillation step was successful. The corresponding coefficients of the Bell-projectors, see Eq.~\eqref{rhosvier}, transform as
\begin{equation}
\label{succmapping}
\begin{pmatrix} 
   \As_{i} \\ 
   \Bs_{i} \\ 
   \Cs_{i} \\ 
   \Ds_{i} 
\end{pmatrix} =
\frac{1}{\probs_i} 
\begin{pmatrix} 
   \big(\As_{i-1}\big)^2 + \big( \Bs_{i-1} \big)^2 \\ 
   2\; \Cs_{i-1} \; \Ds_{i-1} \\ 
   \big( \Cs_{i-1} \big)^2  + \big( \Ds_{i-1} \big)^2  \\ 
   2\; \As_{i-1} \; \Bs_{i-1} 
\end{pmatrix}
\end{equation}
with the corresponding success probability $\probs_i$. This mapping was studied in detail in Ref.~\cite{CMA98}.

If, however, the $i^\text{th}$ step was unsuccessful we have in principle a mapping from $\rhos_{i-1}$ to $\rhou_i$. In general, the density operator $\rhou_i$, see also Eq.~\eqref{rhouvier}, has a lower fidelity than $\frac{1}{2}$, which is the fidelity of the density operator that can be generated by a LOCC process. If the step was unsuccessful we therefore assume from now on that Alice and Bob perform appropriate operations to prepare local qubits with fidelity $\fidu=\frac{1}{2}$.

A full iteration of the entanglement distillation for finite quantum resources has to take into account all possible combinations of these successful and unsuccessful steps. Our aim then is to obtain at the end of a CNOT distillation an entangled qubit pair with an average fidelity as high as possible.  

Note that one can also ask a different question for a small finite sample. Can we distil a single qubit pair with very high fidelity if we allow for a lower success probability? That is, we do not care if the distillation fails several times, but when it is successful it must deliver a highly entangled pair. Then the CNOT distillation process seems to be not very suitable. In this case quantum hashing can turn out to be a powerful method.

\begin{figure}[t]
\begin{center}
\includegraphics{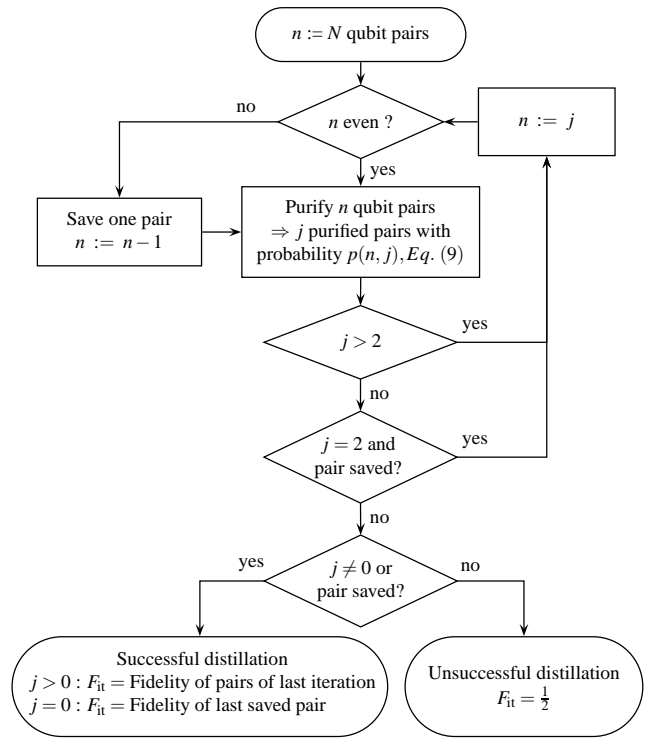}
\end{center}
\caption{\label{CNOTFloat}Flow chart for the iterative CNOT distillation scheme with finite quantum resources. Starting from $N$ qubit pairs we extract a distilled pair. In every step $n/2$ projective measurements are performed and some qubit pairs have to be discarded. To increase the average fidelity a backup pair can be saved in every iteration step for an odd number $n$. Moreover, the iteration stops if only two pairs are left after the projective measurement and no backup pair was saved previously. Rarely the distillation may still fail if no pair was saved and all pairs after the last projection have to be dismissed. The scheme also defines the corresponding fidelity $\Fit$ of the iterative process. Averaging over many runs we obtain $\average{\Fit}(N)$ for $N$ qubit pairs.}   
\end{figure}

The complete iterative CNOT distillation process for finite resources can be expressed in algorithmic form. For a given initial density operator the process depends only on the initial amount of entangled qubit pairs shared by Alice and Bob. An important iterative improvement of the process can be achieved for odd numbers of qubit pairs. In each step of the iteration we may obtain an odd number $n$ of pairs and hence in the simplest case we can store the additional pair as a backup \cite{FMF01}. This backup pair can be used whenever we would have to discard all pairs in some further step of the iteration. 

The stop condition of the algorithm depends on the specific aim of the process. As we have already seen in Eq.~\eqref{aveinzel}, it is impossible to obtain an increase of the average fidelity for two qubit pairs. So if we end up with only two pairs left after a distillation, it seems to be unwise to continue with the iteration, in particular if we have stored no backup so far. The corresponding flow chart of such an algorithm is shown in Fig. \ref{CNOTFloat}. The average fidelity $\average{\Fit}(N)$ of the iterative process can be computed using this algorithm.

In the following example we present the behaviour of such an iterative distillation process. We simulate the process using the probability $p(n,j)$, Eq.~\eqref{pn}, to obtain $j$ qubit pairs when starting from $n$. Moreover, we need the mapping of Eq.~\eqref{succmapping}.

\section{Example of the iterative distillation for small finite sets}

To  demonstrate the behaviour of the iterative distillation we have again chosen Werner-states
($\Bs_0=\Cs_0=\Ds_0=(1-\As_0)/3$) as initial states.

\begin{figure}[ht] 
\begin{center}
\includegraphics{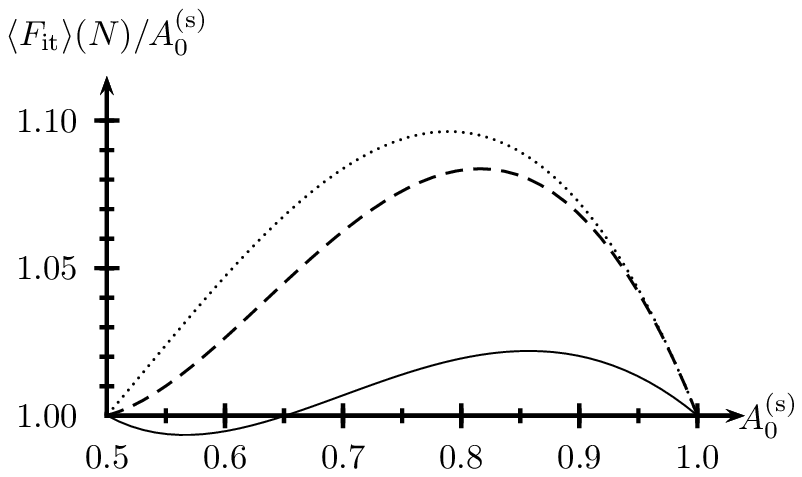}
\end{center}
\caption{\label{FidelityDevelopMulti}Relative average fidelity $\average{\Fit}(N)/\As_0$ of the iteration as a function of the initial fidelity $\As_0$ for different amounts $N$ of qubit pairs. The initial states are again Werner-states ($\Bs_0=\Cs_0=\Ds_0=(1-\As_0)/3$). For $N=4$ qubit pairs (solid line) we only get an improvement if $\As_0 > 0.65$. For greater amounts of qubit pairs we always get an improvement. The advantage of the backup can be seen by comparing the two cases $N=5$ (dotted line) and $N=6$ (dashed line). Due to the fact that for odd numbers we always have a backup pair, the average fidelity becomes higher as compared to the surrounding even cases.}   
\end{figure}
First we compare in Fig.~\ref{FidelityDevelopMulti} the relative average fidelity $\average{\Fit}(N)/\As_0$ as a function of the initial fidelity $\As_0$ for small resources of qubit pairs. In accordance with the algorithm of the previous paragraph we have calculated the final average fidelity $\average{\Fit}(N)$ with backup for $N=4$ (solid line), $N=5$ (dotted line) and $N=6$ (dashed line) pairs. We see that $N=5$ pairs are on average always superior to the nearby even cases, because for an odd $N$ already in the first step of the distillation a backup pair is stored. For $N=4$ there is a minimal initial fidelity which is needed to have a successful iterative distillation.

\begin{figure}[ht]
\begin{center}
\includegraphics{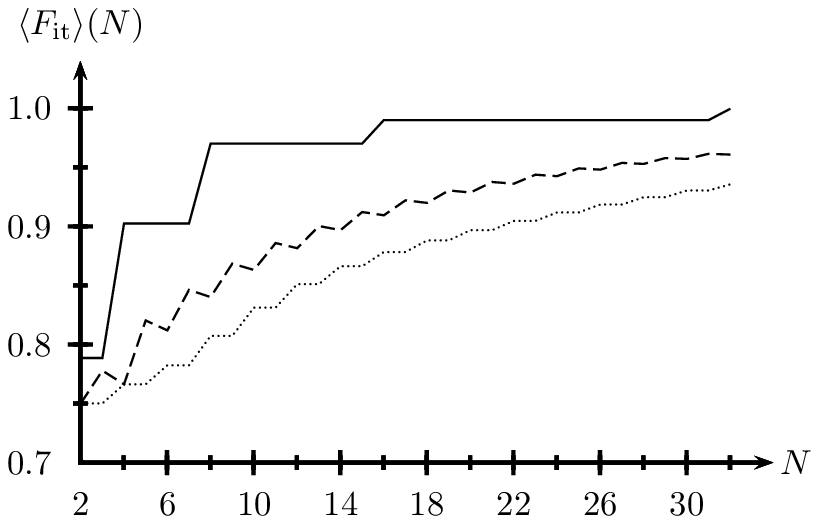}
\end{center}
\caption{\label{CNOTCompare} Comparison of the average fidelities of an iteration depending on the amount $N$ of initially accessible qubit pairs. The plot shows the average fidelities of iterative schemes without backup (dotted curve), with backup (dashed curve), Fig~\ref{CNOTFloat}, and the mapping of the completely successful case (solid curve), Eq.~\eqref{succmapping}. We show their behaviour for the mentioned example state with an initial fidelity $\As_0=0.75$. The plot again shows the advantage of storing one pair in the case of odd numbers $N$. It turns out that for even numbers $N$ it is on average advantageous to drop one pair and to start the distillation with one pair less. This shows that simple backup schemes improve the CNOT distillation and more sophisticated methods can certainly be discussed.}
\end{figure}
Second, we show in Fig.~\ref{CNOTCompare} the generic behaviour of the average fidelity $\average{\Fit}(N)$ depending on the available amount $N$ of pairs. We compare the iterative results without and with backup for a fixed initial fidelity $\As_0=0.75$ to the completely successful mapping, Eq.~\eqref{succmapping}. 
Storing a backup pair in the case of an odd amount clearly leads to a higher average fidelity. The zigzag in the curves reveals the difference between odd and even numbers $N$ of pairs. If the iteration starts with an odd number of entangled qubit pairs, at least one pair is always saved. One can even see that for even numbers it is on average better to drop one pair before performing the first distillation step. Although this means to start with one pair less, the average fidelity is higher. This is most obvious for small finite ensembles.

\section{Conclusions}

In the present paper we have analysed an iterative scheme for a known distillation protocol. This protocol is especially useful for the implementation of an iteration since it puts very low restrictions on the density operators of the processed systems. In particular, we have emphasised the application to finite quantum resources. In this respect it was possible to demonstrate the limitations on the needed number of entangled qubit pairs as well as on their initial entanglement for a successful distillation. Stronger entanglement can be obtained iteratively already for small initial numbers of pairs, even though many pairs have to be sacrificed in order to obtain one distilled pair. We have lowered this loss by introducing backup pairs in our algorithmic description.  

Clearly this is not the only possibility to achieve iterative entanglement distillation. First, one can think of recycling backup systems in the distillation process. This, however, works only in a rather narrow regime in which the fidelities of the recycled pairs are close. Second, it is also possible to iterate completely different distillation methods, like quantum hashing \cite{BBP96} for finite resources. Both possibilities need a considerable amount of book keeping in contrast to the iteration presented here.

Finally, it would be important to simulate distillation for an experimental system, which offers a way to control a finite number of qubit pairs. A possible candidate for this would be an optical lattice filled with atoms, which can be controlled by collisions \cite{BCJ99,JBC99,MGW03}.
\begin{acknowledgments}
We acknowledge financial support by the Deutsche Forschungsgemeinschaft within the \emph{Schwerpunktprogramm Quanten-Informationsverarbeitung} (SPP1078).
\end{acknowledgments}

\end{document}